\documentstyle[epsfig]{mn}
 at 10truept
\title
     [ Bias \& Calibration in Cosmic Shear]
{\vglue-3.0truecm \centerline{\it\small 
For submission to Monthly Notices}
\vglue 2.5truecm
Cosmic Shear Bias and Calibration in Dark Energy Studies
\author
     [ A. N. Taylor \& T. D. Kitching]
     { A. N. Taylor$^1$\thanks{ant@roe.ac.uk} \& T. D. Kitching$^{2}$\\
 	$^1$ Scottish Universities Physics Alliance, 
     Institute for Astronomy,
     School of Physics and Astronomy,
     University of Edinburgh,\\
     \,\,\, Royal Observatory,
     Blackford Hill,
     Edinburgh, EH9 3HJ,
     U.K.\\
     $^2$ Mullard Space Science Laboratory,  
     University College London, 
     Holmbury St. Mary, 
     Dorking, 
     Surrey, 
     RH5 6NT, UK}}
\newcommand{\be}{\begin{equation}}
\newcommand{\ee}{\end{equation}}
\newcommand{\ba}{\begin{eqnarray}}
\newcommand{\ea}{\end{eqnarray}}
\newcommand{\nn}{\nonumber \\}

\newcommand{\thetab}{\mbox{\boldmath $\theta$}}

\newcommand{\de}{\partial}

\newcommand{\lgl}{\langle}
\newcommand{\rgl}{\rangle}

\newcommand{\Tr}{\mbox{\rm Tr}}

\newcommand{\llb}{\mbox{\boldmath $\ell$}}
\newcommand{\ellb}{\mbox{\boldmath $\ell$}}

\newcommand{\x}{\mbox{\boldmath $x$}}

\newcommand{\C}{\mbox{\boldmath $C$}}

\newcommand{\psib}{\mbox{\boldmath $\psi$}}
\newcommand{\0}{\mbox{\boldmath $0$}}

\newcommand{\A}{\mbox{\boldmath $A$}}

\newcommand{\V}{\mbox{\boldmath $V$}}
\newcommand{\Q}{\mbox{\boldmath $Q$}}

\newcommand{\half}{\frac{1}{2}}
\newcommand{\M}{\mbox{$\mathcal{M}$}}

\newcommand{\calP}{\mbox{$\mathcal{P}$}}
\newcommand{\calL}{\mbox{$\mathcal{L}$}}
\newcommand{\Mb}{\mbox{\boldmath $M$}}
\newcommand{\mub}{\mbox{\boldmath $\mu$}}

\newcommand{\calF}{\mbox{$\mathcal{F}$}}

\newcommand{\calA}{\mbox{$\mathcal{A}$}}
\newcommand{\calB}{\mbox{$\mathcal{B}$}}
\newcommand{\calC}{\mbox{$\mathcal{C}$}}

\newcommand{\calH}{\mbox{$\mathcal{H}$}}
\newcommand{\Psib}{\mbox{\boldmath $\Psi$}}
\newcommand{\calAb}{\mbox{\boldmath $\mathcal{A}$}}
\newcommand{\calMb}{\mbox{\boldmath $\mathcal{M}$}}
\newcommand{\Gb}{\mbox{\boldmath $G$}}
\newcommand{\calG}{\mbox{$\mathcal{G}$}}
\newcommand{\calJ}{\mbox{$\mathcal{J}$}}
\newcommand{\calW}{\mbox{$\mathcal{W}$}}

\newcommand{\calFb}{\mbox{\boldmath $\mathcal{F}$}}
\newcommand{\calBb}{\mbox{\boldmath $\mathcal{B}$}}

\newcommand{\sinc}{\mbox{$\rm{sinc}$}}

\begin{document}

\maketitle

\begin{abstract} 
With the advent of large-scale weak lensing surveys there is a need to understand how realistic, scale-dependent systematics bias cosmic shear and dark energy measurements, and how they can be removed.
Here we describe how spatial variations in the amplitude and orientation of realistic image distortions convolve with  the measured shear field, mixing the even-parity convergence and odd-parity modes, and bias the shear power spectrum. 
Many of these biases can be removed by calibration to external data, the survey itself, or by modelling in simulations.
The uncertainty in the calibration must be marginalised over and we calculate how this 
 propagates into parameter estimation, degrading the dark energy Figure-of-Merit. 
We find that noise-like biases affect dark energy measurements the most, while spikes in the bias power have the least impact, reflecting their
 correlation  with the effect of cosmological parameters.
We argue that in order to remove systematic biases in cosmic shear surveys and maintain statistical power  effort should be put into improving the accuracy of the bias calibration rather than minimising the size of the bias. In general, this appears to be a weaker condition for bias removal. 
We also investigate how to minimise the size of the calibration set for a fixed reduction in the Figure-of-Merit. These results can be used to model the effect of biases and calibration on a cosmic shear survey accurately, assess their impact  on the measurement of modified gravity and dark energy models, and to optimise surveys and calibration requirements.
\end{abstract}


\section{Introduction}


The observed accelerated expansion of the Universe presents cosmology with one of its biggest challenges. While this acceleration can be accommodated by the inclusion of a classical cosmological constant, quantum corrections from vacuum fluctuations are uncontrolled, leading to runaway values which exceed the observed energy-density by many orders of magnitude. 
A compelling fundamental solution is so-far elusive, but cosmologists have proposed a large number of alternative low-energy effective theories, called dark energy models if they inhabit  the matter sector and modified gravity if they are in the gravity sector, which aim to cast light on these new forces. 
While solving the cosmic acceleration problem, such theories alter the growth of structure in the Universe leaving traces of these new forces
which may be detectable in galaxy clustering and weak lensing surveys. High accuracy observations are now needed to test the subtle differences in these dark energy and modified gravity theories to constrain the wide range of possibilities and to point the way to a more fundamental theory. Such high accuracies come from large-scale ground and space-based surveys, which will provide statistical accuracy but will be be limited by their systematic biases
(e.g., VST-KiDS\footnote{http://www.astro-wise.org/projects/KIDS}, DES\footnote{http://www.darkenergysurvey.org}, HSC\footnote{http://www.naoj.org/Projects/HSC/index.html}, LSST\footnote{http://www.lsst.org}, {\em Euclid}\footnote{ http://www.euclid-ec.org, Laureijs, et al. (2011),}, {\em WFIRST}\footnote{http://www.stsci.edu/wfirst/}).
For these surveys to be successful, systematic biases should be controlled to an unprecedented level -- to within the bounds set by the statistical uncertainty. The origin of these biases, and the accuracy to which they can be removed, will need to be studied at every step of the data analysis,  from observation to parameter estimation. 


In order to develop the required technology for surveys and data analysis,  a rapid and accurate method is needed to assess the impact of systematic biases  on cosmological studies and to determine which biases have the most potential to damage dark energy studies. In particular, dark energy and modified gravity models generically introduce scale-dependent deviations from $\Lambda$CDM structure formation  which could have similarities to the systematic effects on different scales. 
The development of studies of the effect of systematic biases on weak lensing power spectra studies has developed over the last decade (see for example:  Ishak et al. 2005; Knox et al. 2006; Bernstein 2006; Huterer et al. 2006; Taylor et al. 2007;  Kitching, Taylor \& Heavens 2008; Amara \& R{\'e}fr{\'e}gier 2008; Paulin-Henriksson et al. 2008;  Kitching et al. 2009; Bernstein \& Huterer 2010; Kitching et al. 2012;  Massey et al. 2013; Cropper et al. 2013; Cardone et al. 2013; Kitching et al. 2015)
The effect of scale-dependent systematic biases on cosmic shear power was first discussed by Huterer et al. (2006) who studied the impact of scale-dependent additive image distortions and a constant multiplicative bias  on a range of cosmological parameters, along with the effect of biases in photometric redshifts. 
 Amara \& R{\'e}fr{\'e}gier (2008)  also studied the bias due to image distortions arising from a scale-dependent additive bias, exploring a number of functional forms for the scale-dependence,  and a constant multiplicative bias. They also investigated the effect of redshift-dependence in these biases. 
These studies found that a constant multiplicative bias on shear should be kept below $\sim 10^{-3}$ and any additive bias shear power should be below  $\sim 10^{-10}$, to prevent the bias dominating over noise.

Given that the cause of systematics was still poorly understood, Kitching et al. (2009) argued that instead of choosing a fixed functional form for systematics, one should average over all possible functional forms in a Monte-Carlo approach.
Building on the work of Paulin-Henriksson et al. (2008), who studied how constant biases in the Point Spread Function (PSF) affect the measurement of shear,  Massey et al. (2013) showed how inaccuracies in the PSF,  the measurement of galaxy shapes and weighting, and the effect of Charge Transfer Inefficiency (CTI)  propagate into the bias and error on the estimated shear. 
Using the shear power spectrum bias formalism introduced by Kitching et al. (2012) in the GREAT10 Challenge, with both scale-dependent multiplicative and additive shear power bias terms, 
Massey et al. (2013) averaged over all functional forms for the scale-dependence of the biases and, based on their effects on the dark energy Figure-of-Merit, constrained their amplitude for space-based weak lensing surveys.
Cropper et al. (2013) subsequently took these constraints and propagated  them back to constraints on individual sources of bias in shear measurement, assuming the bias and uncertainties  were independent of scale. 

In order to better understand the source of  scale-dependency in image distortions,  Kitching et al. (2015) simulated systematics in the PSF, CTI and shear estimation, and measured their effect on the shear power spectrum. 
Using the multiplicative and additive bias power formalism, they propagated this into the bias and the covariance  on measured cosmological parameters and found that a survey  could minimise the impact of bias by randomising the observing strategy so that the systematic power became noise-like. They also investigated the removal of sharp spikes in the shear power  due to discontinuities on CCD and field-of-view scales. 

From the Kitching et al. (2015) study it became clear that the multiplicative and additive biases of the shear field are expected to be spatially variable and scale-dependent. However, spatial variations in the the shear distortion on the sky correspond to a convolution of the shear signal with the systematic bias in the Fourier domain, rather than a simple multiplicative factor. Hence the shear power spectrum will be convolved with the bias power, and so improved modelling is needed. 
 In addition, studies to-date have derived constraints on the size of any image bias effects by comparing the bias in the final cosmological parameter, or Figures-of-Merit, with the expected random error.  This assumes that new algorithms  can be developed which will mitigate these biases to the level required. However, in many cases the biases may be too complex to accurately model, or the modelling may be too slow for practical application. An alternative is that the bias is removed by calibration with external data, or by the survey itself, or through modelling in simulations. In this case the relevant factor is the accuracy to which the calibration can be carried out, and how this error propagates into dark energy studies.

In this paper we address these issues by developing a  formalism to study how spatially varying  systematics arising from image distortions affect the shear and the inferred convergence field (Section 2), and propagate them into the cosmic shear power spectrum (Section 3). We study how realistic biases in the shear power spectrum can be removed by calibration, and the resulting uncertainty marginalised over. The effect of this on parameter estimation and the impact on the dark energy Figure of Merit is explored (Section 4). Taking realistic examples of image distortions, we show how to optimise the constraints on the amplitudes of a set of systematic effects for a given shear survey, to minimise the effect on cosmological parameters (Section 5).


\section{Weak Lensing Bias}


The response of a measurement of cosmic shear, $\widehat{\gamma}$,  to the true shear field, $\gamma$, which is of order a few percent, in the presence of image distortions can be characterised by a linear model with a local multiplicative factor, $m$, an additive term, $c$ (Heymans et al. 2006, Massey et al. 2007),  and a non-local convolution term, $h$, such that 
 \be
 	\widehat{\gamma} = (1+m) \gamma + h * \gamma + c.
 \ee
 The multiplicative factor, $m$, is a spin-2 field representing a change in amplitude and a local rotation of the shear field, while the additive term is an arbitrary spin-2 shear-like distortion. A multiplicative bias can arise due to miscalibration of the shear measurement caused, for example, by incorrect modelling of the ellipticity or size of the PSF, residual CTI and noise-bias or shear estimation effects. 
 The spin-2 multiplication bias can be written 
$
	m(\thetab) 
	= m_0(\thetab) e^{i 2 \phi_{\small m}({\small \thetab})},
$
(Kitching, Taylor \& Heavens 2008)
where $\phi_m(\thetab)$ is the local rotation of the phase and $m_0(\thetab)$ is a local, scalar modulation of the shear amplitude. Massey et al. (2007) and Kitching et al. (2012) also investigated a quadratic distortion term, $(1+q) \gamma^2$, but found that second order terms were negligible.  
 The additive bias, $c$, can arise due to systematics in the ellipticity and shape of the PSF, or from CTI leaving residual streaks in the image. 
 The convolution term, $h$, represents a distortion which depends on the shear field at other positions, which may arise due to close packing or blending of galaxy images, and is again a spin-2 field.

Fourier transforming the measured shear field on a flat sky, we find
  \be
  \label{eq:2}
 	\widehat{\gamma}(\llb) =  \gamma(\ellb) +
		 \int \! \frac{d^2 \ellb'}{(2\pi)^2}\,  m (\ellb-\ellb') \gamma(\ellb')    + c(\ellb),
 \ee
 where the spatially varying multiplicative bias on the sky now convolves the shear field. For simplicity we have absorbed the convolution distortion, $h$, into the multiplicative bias, $m$. The shear field can also be decomposed on the full curved sky in spherical harmonics (see, for example Brown, Castro \& Taylor, 2005), but for simplicity we use a flat-sky approximation here.
The shear signal can then be decomposed into even-parity  convergence  modes, $\kappa$, and odd-parity $\beta$-modes\footnote{From here on we shall refer to $B$-modes in lensing as $\beta$-modes, where $E$-modes correspond to the convergence field, $\kappa$.} by a rotation of the shear in the Fourier domain,
 $
 	\kappa(\ellb) + i \beta(\ellb) = e^{-2i \varphi_\ell} \gamma(\ellb),
 $
 where $\varphi_\ell$ is the angle between the wavevector, $\ellb$, and an arbitrary axis on the sky. Assuming only a scalar multiplicative bias, $m_0$,
the measured $\kappa$ and $\beta$ modes are distorted by
\ba
\label{eq:3}
	\Delta \kappa(\ellb) \!\!\!\!\!&=&\!\!\!\!\!
		\int \!\! \frac{d^2 \ell'}{(2\pi)^2}\, \!
		m_0(\ellb-\ellb')  \left(\kappa(\ellb') \cos 2 \varphi_{\ell\ell'}   -
		  \beta(\ellb')	\sin 2 \varphi_{\ell\ell'} \right)
		  \nn &&
		  \!\!\!\!\!+  \, c_\kappa(\ellb),  
		  \\ 
	\label{eq:4}
		  	\Delta \beta(\ellb) \!\!\!\!\!&=&\!\!\!\!\!
		\int \!\! \frac{d^2 \ell'}{(2\pi)^2}\, \!
		m_0(\ellb-\ellb')  \left(\beta(\ellb') \cos 2 \varphi_{\ell\ell'} +
		 \kappa(\ellb') \sin 2 \varphi_{\ell\ell'}  \right)
		  \nn &&
		  \!\!\!\!\!+ \, c_\beta(\ellb), 
\ea
where $\Delta \kappa = \widehat{\kappa} - \kappa$ and  $\Delta \beta = \widehat{\beta} - \beta$ are the changes in the convergence and $\beta$-fields, and $\varphi_{\ell\ell'}=\varphi_\ell - \varphi_{\ell'}$ is the angle between the Fourier modes.  We have also decomposed the additive bias into even and odd parity modes, where $c_\kappa(\ellb) + i c_\beta (\ellb)  = e^{-2i \varphi_\ell}c(\ellb) $. The mode-mixing effect of the full spin-2 multiplicative bias is slightly more complex and we present full expressions  in Appendix A.
 As discussed by Kitching et al. (2012), the effect of a spatially varying multiplicative bias is similar to that of a survey mask in the Pseudo-Cl (PCL)  power spectrum formalism (Hivon et al., 2002)  for CMB polarisation (Brown, Castro \& Taylor, 2005), and so we can easily generalise our results to a masked survey. 

\section{Cosmic Shear Power}

The correlations of the different Fourier modes of the shear fields, $(X,Y)=(\kappa,\beta)$, for different $\ell$-modes is given by 
\be
\label{eq:5}
	\lgl  X(\llb) Y^*\!(\llb')\rgl  = (2 \pi)^2 C^{XY}\!(\ell) \,  \delta_D(\llb-\llb'),
\ee
where  $C^{XY}\!(\ell)$ is the convergence, $\beta$-mode and cross-power spectrum. We assume all fields are statistically homogeneous and isotropic on a flat sky, and $\delta_D(\ellb)$ is the Dirac delta function. 
The measured convergence power spectrum on a flat, finite patch of sky of area, $A$, is given by
\be
	\widehat{C}^{\kappa\kappa}\!(\ell)=
	\frac{1}{A} \left\lgl | \widehat{\kappa}(\ellb) |^2 \right\rgl ,
\ee
where we have approximated the zero-lag delta function by $ \delta_D(\0) = A/ (2 \pi)^2  $. 

We denote the systematic bias fields  on the sky by $Z(\thetab)=(m,c_\kappa,c_\beta)$ for  each of the multiplicative/convolution fields and the even and odd parity modes of the additive biases. These biases can be split into a constant term across the survey, $Z_0 = b_Z$, a spatially varying deterministic bias,
$
	\Delta Z(\thetab),
$
around the mean which can arise from variations which can be modelled by a template,
and a stochastic term, $\delta Z(\thetab)$, that  arise from either noise in the measurement of the bias or other indeterminate aspects of the bias that can only be modelled statistically (e.g., Massey et al., 2013); hence
$	
	Z = b_Z + \Delta Z  + \delta Z.
$
The correlations of the Fourier modes of the fluctuating part of the bias are given by 
\be
\label{eq:7}
		\lgl \delta Z(\llb)  \, \delta Z^{\!*}\!(\llb') \rgl = (2 \pi)^2 \calC_Z(\ellb) \delta_D(\llb-\llb'),
\ee
where $\calC_Z(\ellb)$ is the power spectrum of the bias fluctuations.
The assumption of statistical isotropy can be relaxed to allow for anisotropic  effects such as, for example, from CTI or other effects aligning with the CCD pixels, and with other directional dependences.

Taking equations (\ref{eq:3}) and (\ref{eq:4}), and using equations (\ref{eq:5}) and (\ref{eq:7}), we can calculate the correlators of the measured convergence and $\beta$-modes. We present the full correlations of the observed Fourier modes of the convergence and $\beta$ fields for an arbitrary spin-2 multiplicative bias  in Appendix \ref{Correlations of the observed Fourier modes}, equations (\ref{eq:conv_FFT_corr}) to (\ref{eq:conv_FFT_corr2}), from which we see that the observed convergence power  is
\be
\label{eq:obs_conv}
	\widehat{C}^{\kappa\kappa}\!(\ell) =
	 \left(1 +
	b_{m} \right)^2 C^{\kappa\kappa}\!(\ell) 
	 +
	 \int \! \frac{d^2 \ell'}{(2\pi)^2} 
	\calG (\ellb-\ellb') 
	 C^{\kappa\kappa}\!(\ell') 
	 	 +  \calA^{\kappa\kappa}\!(\ell), 
\ee
where $b_m$ is a constant multiplicative bias,
the convolution  kernel is
\be	
\label{eq:kernel}
	\calG(\ellb-\ellb') =  
	  \left( \frac{1}{ A} 
	 |\Delta b_m(\ellb-\ellb')|^2
	 +
	\calC_{m}(\ellb-\ellb') \right) \cos^2 2 \varphi_{\ell \ell'},
\ee
and the additive bias is
$
	\calA^{\kappa\kappa}\!(\ell)  =  
	  |\Delta b_{c_\kappa}(\ellb)|^2/A + \calC_{c_\kappa}(\ell).
$
Any constant additive bias appears only in the $\ell=0$ mode and can be ignored.
Similar expressions for the $\beta$-mode power spectrum are derived in Appendix \ref{B-mode power spectra}. From equation (\ref{eq:obs_conv}) we see that the constant bias over the survey, $b_m$, factors out into a multiplicative bias of the convergence power spectrum. The scale-dependence of the multiplicative bias, $\Delta b_m(\ellb)$, comes in only through the convolution term at second-order, and is weighted by the inverse survey area. Finally, the power spectrum of the indeterminate stochastic bias, $\calC_m(\ellb)$, is also convolved with the convergence power spectrum, while the additive power is composed of the square of the additive shear bias variation and its power spectrum. 

Following Kitching et al. (2015), we can model the effect of realistic bias power spectra.
If the bias power spectrum is a sharp peak at $\ell=\ell_0$, we can approximate it by a delta-function,  $\calC_{m}(\ell)= (2\pi)^2 \calC_0 \delta_D(|\ellb-\ellb_0|)/\ell$. 
The change in the measured shear power is then $\Delta \widehat{C}^{\kappa\kappa}\!(\ell) = \calC_0 \, C^{\kappa\kappa}(|\ellb-\ellb_0|)$, which shifts the shear power origin to $\ell_0$, reflects about it and rescales the amplitude. 
A  spike of power centred at wave vector $\ellb=\ellb_0$ would not induce $\beta$-modes, but here we have assumed an isotropic distribution localised about a single wavelength which will generate curl modes of similar amplitude. 
For a constant noise-like bias power, $\calC_{m}(\ell) = C_{\rm N}$, the change in the shear power is $\Delta \widehat{C}^{\kappa\kappa}_\ell =   C_{\rm N} \sigma^2_\kappa /2$, where the other half of the convolved noise power has been converted into a $\beta$-mode power spectrum (see Appendix \ref{B-mode power spectra}). Kitching et al. (2015) also found that sharp features in a weak lensing survey associated with biases on particular scale, $\theta$,  such as the field of view, resulted in a sinc-like scale-dependent bias power, $C_{\rm S}(\ell \theta) \propto \sinc (\ell \theta)$. Since this is noise-like on large-scales, to a good approximation the change in the shear power is $\Delta \widehat{C}^{\kappa\kappa}\!(\ell) =   C_{\rm S}(\ell \theta) \, \sigma^2_\kappa /2$, where again a $\beta$-mode power of equal amplitude is generated. 

\begin{figure*}
  \vspace{-0.4cm}
   \includegraphics[width=2.3\columnwidth]{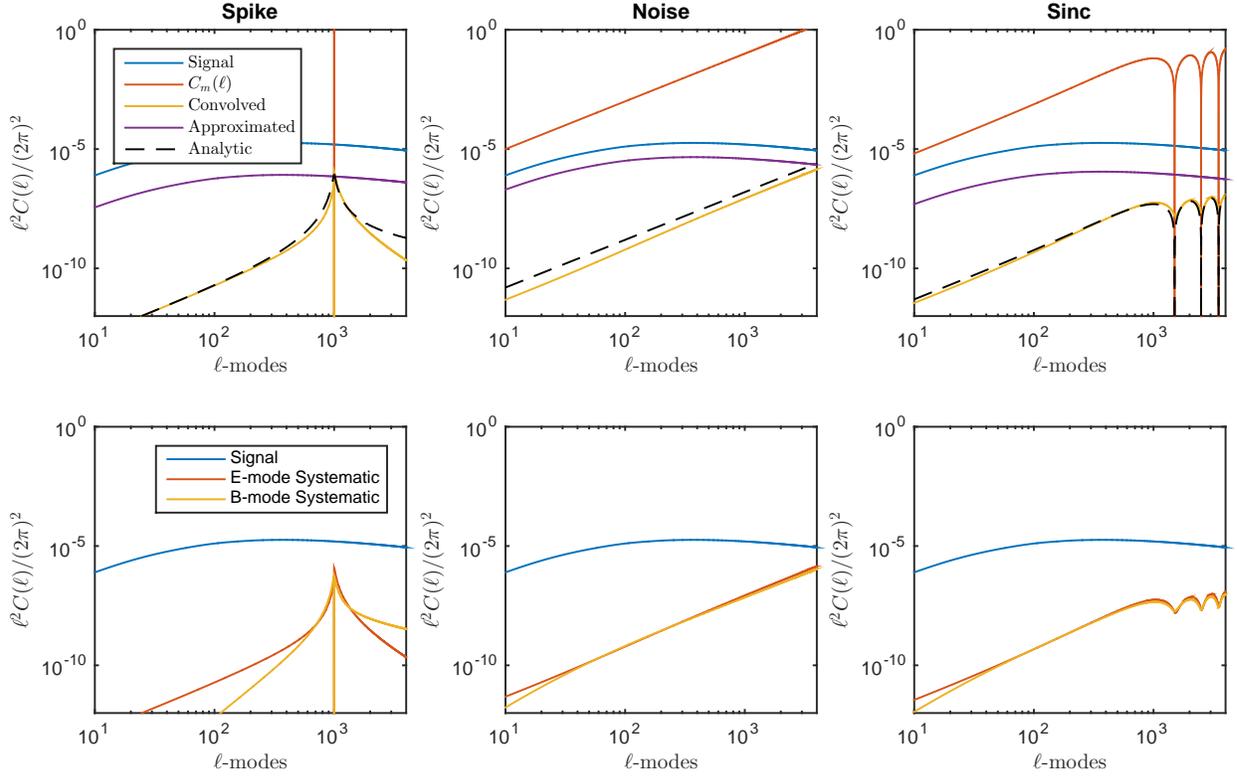}
     \vspace{-0.5cm}
  \caption{\em Convolution shear-convergence and $\beta$-mode power bias. In all panels the blue lines are the true convergence power spectrum signal, $C^{\kappa\kappa}_\ell$;  in the top panels the orange lines are the multiplicative noise power, $\calC_m(\ell)$; yellow lines are the bias power numerically convolved with the convergence power, while the black dotted lines are the exact analytic convolution from Section 3, which are in good agreement with the numerical convolutions. Purple lines are the approximate convolution given by equation (\ref{eq:rapidvarV}), which provides a poor approximation. In the lower plots the orange lines are again the bias power spectra and yellow lines are the $\beta$-mode power generated by mode mixing.
   {\bf Left Panels:} The convergence power is convolved with a sharp bias feature at $\ell =10^3 $, leading to an extended bias power, with good agreement between numerical and analytic convolutions. $\beta$-mode power is generated from power at $\ell_0$ (lower plot); 
   {\bf Middle Panels:} A constant noise-like bias leads to a constant convolution bias, suppressed by the shear variance, as predicted by the analytic results. A constant $\beta$-mode power spectrum is generated with the same amplitude as the bias convergence power (lower panel); 
   {\bf Right Panels:} A sinc-like bias power is convolved with the convergence power, where the shape and amplitude is well approximated by the analytic model of Section 3. We find a $\beta$-mode power of equal shape and amplitude (lower panel).
   In all cases the convolved convergence and bias powers are
  poorly modelled by the multiplicative bias approximation of equation (\ref{eq:rapidvarV}).
    }
    \label{fig:1}
\end{figure*}

Figure \ref{fig:1} shows the shear convergence power for a $\Lambda$CDM model with parameters taken from Planck (Planck 2015, XIII) along with the three generic, multiplicative scale-dependent biases:
a localised spike in bias power with $\calC_0=1$, which may arise from residual corrections on a particular scale (Figure \ref{fig:1}, left panels); a constant noise-like power bias  with  $C_N=10^{-10}$, which may arise from random variations in a residual bias across the survey on all scales (Figure \ref{fig:1}, middle panels); and a sinc-like bias function (Figure \ref{fig:1}, right panels) of the form $C_S(\ell \theta_s) = 10^{-10} \sinc( \ell \theta_s)$, where $\theta_s =10^{-3} {\rm rad}$, which may arise due to a residual bias over a finite patch such as the CCD of field-of-view scales (Kitching et al. 2015). The convolution of each of these multiplicative biases with the shear power is also shown in each panel (yellow lines), along with the analytic expressions found above (black, dashed lines). In all cases we find that the analytic formula predict the numerical data well.

 As expected the local bias spike leads to a bias spread out in $\ell$-space, although still highly peaked and centred around the spike, while the noise-like bias power convolution gives rise to a constant bias whose amplitude is suppressed by half of the shear-convergence variance, $\sigma^2_\kappa/2$.  The convolved sinc-like power has the same form as the bias power, also suppressed by a factor of half the shear-convergence variance. There is an induced $\beta$-mode power for the sharp spike (Figure \ref{fig:1}, lower left panel), while the $\beta$-mode noise power is equal to the convergence noise power (Figure \ref{fig:1}, lower middle panel). We also find the $\beta$-mode power from the sinc-bias is the same as the convergence power (Figure \ref{fig:1}, lower right panel), as predicted.

A commonly-used model for bias in the shear power spectra is to assume a scale-dependent multiplicative term (Kitching et al. 2012, Massey et al. 2013, Kitching et al. 2015), where 
\be
\label{eq:Chat}
	\widehat{C}^{\kappa\kappa}\!(\ell)  = \big[1+\M(\ell)\big] C^{\kappa\kappa}\!(\ell) + \calA^{\kappa\kappa}\!(\ell).
\ee
As we have argued, any local fluctuations in the calibration of the shear signal will lead to a convolution of the shear power or, if the bias is a constant, a constant multiplicative term. Only non-local spatial distortions will lead to a pure multiplicative term. The multiplicative shear power bias is likely to have a different response to a convolved shear power. 
To see if we can approximate the convolution bias by a multiplicative bias we can  assume the shear power varies less than the systematic power and, approximating the shear power by a constant, take it outside the convolution.
If the bias power is isotropic we can carry out the angular integration over the cosine-squared term to give a factor of $\pi$, while integrating over the radial $\ell$-modes yields 
\be
\label{eq:rapidvarV}
		\widehat{C}^{\kappa\kappa}\!(\ell) =
	 \left((1 +  b_m )^2
	+ 
	\half \sigma^2_m \right) C^{\kappa\kappa}\!(\ell) 
	 	 	 +  \calA^{\kappa\kappa}\!(\ell) ,
\ee
where
$
	\sigma^2_m 
$
is the variance of the spatially varying  multiplicative bias. In this limit the convolution by the spatially varying multiplicative bias becomes a constant multiplicative bias factor, $\sigma^2_m/2$, in the shear power. The other half of the variance contributes to the odd-parity $\beta$-mode power (see Appendix \ref{B-mode power spectra}, equation \ref{eq:Cbb}). 
In this limit, we expect the multiplicative bias has only a weak scale dependence and that the main effect is to boost the amplitude of the shear power. In Figure \ref{fig:1} we compare this approximation (purple curves in the upper panels) with the spike, constant and sinc-like bias model power spectra, and find that the approximation is poor. The shape of the convolved shear bias is not well reproduced, tending to over-predict the amplitude of the bias on all scales and incorrectly predict the scale dependence. Hence, we advocate that modelling of the scale-dependence of a spatially varying multiplicative shear bias in the shear power uses  our convolution model rather than a scale-dependent multiplicative term.

\section{Calibration and Removal of Cosmic Shear Bias }

In previous studies the effect of systematics  in cosmic shear measurements has focussed on propagating the biases into the dark energy parameters and setting constraints  such that either the biases are less than the measurement error (e.g. Amara \& R{\'e}fr{\'e}gier 2008), or that the Dark Energy Figure of Merit (DEFoM) is kept above some fixed value (e.g. Massey et al. 2013). This is useful if there is an algorithmic way to remove these biases.
However, in practice many systematics may be too complex  to model to sufficient accuracy, and so they need to be removed by calibration to external data or simulations and the uncertainty on the calibration then marginalised over. This suggests that it is not sufficient  to know how the bias affects the dark energy measurement - we also need to know how marginalisation over the  uncertainty in the calibration  propagates into the measurement.

We can explore the effect of calibration and marginalisation using the Fisher matrix formalism (e.g., Tegmark, Taylor \& Heavens 1997).  Let us assume the 
measured shear convergence power, $\widehat{C}^{\kappa\kappa}_\ell(\thetab)$, given by equations (\ref{eq:obs_conv}) and (\ref{eq:kernel}), is Gaussian distributed, $\calP_m \big(\widehat{C}^{\kappa\kappa}_\ell | \thetab\big)$, and depends on a set of  cosmological parameters, $\thetab$, whose likelihood function, $\calL(\thetab)=\big(\widehat{C}^{\kappa\kappa}_\ell | \thetab\big)$, is also Gaussian distributed in parameter space. The expected cosmological parameter covariance matrix for this likelihood is $C = \lgl \Delta \thetab \Delta \thetab^t\rgl = \calF^{-1}$,
where 
\be
	\calF_{\alpha\beta}  = \frac{4 \pi f_{\rm sky} }{2}
	\int \!\frac{\ell d \ell}{2 \pi} \,  
		\big[\widehat{C}^{\kappa\kappa}\!(\ell) + N\big]^{-2}  
	\frac{\de \widehat{C}^{\kappa\kappa}\!(\ell)}{\de \theta_\alpha}  \,  \frac{\de \widehat{C}^{\kappa\kappa}\!(\ell)}{\de \theta_\beta}  , 
\ee
is the Fisher matrix, $A=4 \pi f_{\rm sky}$ is the area of the survey, and  $N =2 \pi f_{\rm sky}\sigma^2_e / N_{\rm g}$ is the shear noise power for $N_{\rm g}$ galaxies with intrinsic ellipticity dispersion $\sigma_e$.

We assume the convolved and additive bias power are functions of a set of bias parameters, $\psib$, so that $\widehat{C}^{\kappa\kappa}_\ell(\thetab,\psib)$ now depends on $\psib$ and has the distribution $\calP_m \big(\widehat{C}^{\kappa\kappa}_\ell | \thetab, \psib\big)$. These bias parameters can be estimated from external data or simulations, with the distribution  
 $\calP(\psib|\psib_0,\widehat{C}_{\psi\psi'})$, with mean given by the true bias values, $\lgl \psib \rgl = \psib_0$, and covariance matrix, $\widehat{C}_{\psi\psi'}$. We assume this distribution is also Gaussian.
 The biased shear power distribution can now be corrected by marginalising over the calibration measurement distribution, 
 \be	
	\calP_m \big(\widehat{C}^{\kappa\kappa}_\ell | \thetab, \psib_0,\widehat{C}_{\psi\psi'} \big) \!=\!\!
	\int d \psi   \, \calP_m \big(\widehat{C}^{\kappa\kappa}_\ell | \thetab, \psib\big) \calP\big(\psib|\psib_0,\widehat{C}_{\psi\psi'}\big),
\ee
which will correct the bias and widen the shear likelihood.  

Expanding the observed shear-convergence power to first order in the bias parameters, we can carry this marginalisation out analytically (Taylor \& Kitching 2010).
The cosmological parameter covariance matrix from this marginalised likelihood can be found from  the inverse of the marginalised Fisher matrix, $C^M = [\calF^M]^{-1}$, where the marginalised  Fisher matrix is given by the Schur compliment of the cosmological and bias parameter Fisher matrix  (e.g., Taylor \& Kitching 2010),
\be
\label{eq:marg}
	\calF^M_{\theta\theta'} = \calF_{\theta\theta'} - \calF_{\theta\psi} \big[\calF_{\psi\psi'}+\widehat{C}^{-1}_{\psi\psi'}\big]^{-1} \calF_{\psi'\theta'}.
\ee
As the accuracy of the external calibration increases, the second term vanishes and the  parameter variance is unchanged.
However, even if the external calibration is removed the loss of accuracy in the parameters is finite, because we can self-calibrate the biases using the cosmic shear survey itself.

As well as the effect on the cosmological parameter covariance matrix, we can also estimate the effect on the Dark Energy Figure of Merit.
The DEFoM is defined as the inverse area of the $68.3\%$ confidence region of the dark energy  2-parameter space, $w=(w_0, w_a)$, after marginalising over all other cosmological parameters (Albrecht et al. 2006). For Gaussian distributed parameters this is given by the determinant of the dark energy Fisher matrix,
\be
	{\rm F}_{\rm oM}^{\rm DE} = \det \calF^{DE}_{w w'} ,
	\label{eq:DEFoM}
\ee
where $\calF^{DE}_{w w'}$ is the dark energy parameter Fisher matrix found by marginalising the cosmological parameter space over all other cosmological parameters. This is also given by the Schur complement of the full parameter Fisher matrix,
\be
	\calF^{DE}_{w w'} = \calF_{ww'} - \calF_{w \theta} \calF_{\theta \theta'}^{-1} \calF_{\theta' w'},
	\label{eq:DEFisher}
\ee
where, in this expression, $\theta$ are all the cosmological parameters excluding the dark energy $w$-vector.
The effect of calibration and marginalisation on the DEFoM can be calculated by  replacing the Fisher matrices in equation (\ref{eq:DEFisher}) with marginalised ones from equation (\ref{eq:marg}).

We can estimate the effect of calibration on the DEFoM. The fractional change in the DEFoM from a change in the Fisher matrix, to first order, is
\be
	\Delta \ln {\rm F}_{\rm oM}^{\rm DE} =  \Tr \, \left( \Delta \calF^{DE} [\calF^{\rm DE}]^{-1} \right),
\ee
where
$
	\Delta \calF^{DE}_{w w'} =  \Delta \calF_{ww'} - \Delta [\calF_{w \theta} \calF_{\theta \theta'}^{-1} \calF_{\theta' w'}].
$ 
Following marginalisation over the calibration parameters, $\psib$, using equation (\ref{eq:marg}), the fractional change in the DEFoM is 
\be	
\label{eq:DDEFOM}
	\Delta \ln {\rm F}_{\rm oM}^{\rm DE} = - \big(\calF^M_{\psi w} \calF^{-1}_{ww'} \calF^M_{w' \psi'}\big)
	\big[  \calF_{\psi'\psi} +\widehat{C}^{-1}_{\psi' \psi} \big]^{-1},
\ee
where $ \calF^M_{\psi w} = \calF_{\psi w} - \calF_{\psi \theta} \calF^{-1}_{ \theta \theta'} \calF_{\theta' w}$ is the joint  Fisher matrix of dark energy and calibration parameters, marginalising over all other cosmological parameters. Equation (\ref{eq:DDEFOM}) shows explicitly the relationship between the calibration accuracy and degradation of the DEFoM. Again, if the accuracy of external calibration is high the DEFoM is unchanged, while if it is removed, self-calibration limits the reduction  in the DEFoM.

\begin{figure}
  \centering
  \vspace{-0.2cm}
   \includegraphics[width=1.08\columnwidth]{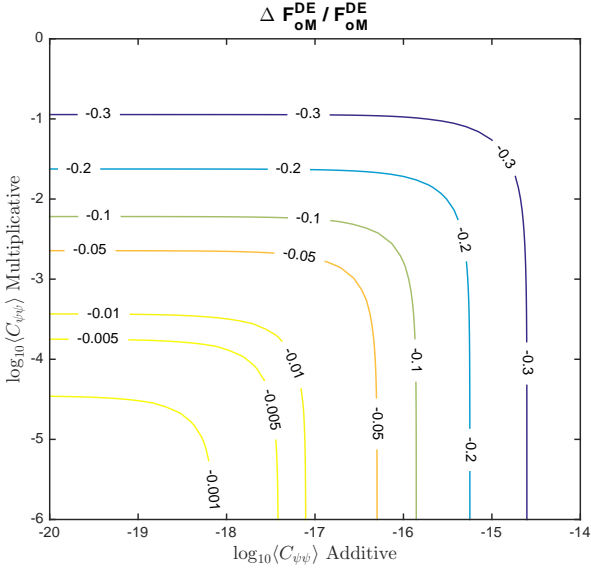}
  \caption{\em   
  The fractional decrease in the dark energy Figure-of-Merit (DEFoM), $\Delta \ln F^{\rm DE}_{\rm oM}$ as a function of the error on the external calibration of an additive bias and multiplicative (convolution) bias, owing to calibration and marginalisation. The functional form of the bias is the sum of the spike, noise and sinc-functions shown in Figure \ref{fig:1}, and we assume only the amplitude requires calibration. The axes are the external calibration variance, $\widehat{C}_{\psi\psi}$, where here the indices $\psi$ and are the amplitude of the multiplicative (convolutions) and additive biases.  Beyond $\Delta \ln F^{\rm DE}_{\rm oM} = -0.3 $ the surface flattens as we reach the self-calibration regime.
    }
    \label{fig:3}
\end{figure}

In dark energy studies we usually want the absolute contribution from the fractional  DEFoM bias to be less than some threshold, $\nu$, so that
 \be
  \label{eq:req_B}
 	|\Delta \ln {\rm F}_{\rm oM} | \le \nu .
 \ee 
As an example, if we consider a constant dark energy equation of state parameter, $w=w_0$,  and one other cosmological parameter, $\theta$, and a single, constant multiplicative calibration parameter, $\psi={\M} = 2 b_m+b_m^2$, with covariance $\widehat{C}_{\small \M\M'} = \widehat{\sigma}^2_{\small \M}  I_{\small \M\M'}$,  the fractional decrease in the DEFoM is 
\be
\label{eq:DEFOMM}
	\Delta \ln {\rm F}_{\rm oM}^{\rm DE} = -\left|\Delta \mu _{\!{\small \M} w_0 }\right|^2 
	\left(\frac{\widehat{\sigma}^2_{\small \M}}{  \widehat{\sigma}^2_{\small \M} +    \sigma^2_{\small \M}} \right),
\ee
where
$ 
\Delta \mu_{\alpha\beta} =\mu_{\alpha \beta} - \mu_{\alpha \theta} \, \mu_{\theta \beta} 
$,
with the implied summation over all other parameters, $\theta$;
$\mu_{\alpha\beta}  = \calF_{\alpha \beta} / \sqrt{\calF_{\alpha\alpha} \calF_{\beta\beta}}$ is the Fisher matrix  correlation coefficient; $\sigma^2_{\small \M}= 1/\calF_{\small \M\M}$ is the self-calibration variance of $\M$ measured from the survey itself; and $\widehat{\sigma}^2_{\small \M}$ is the external calibration variance. 
The decrease in the DEFoM vanishes as the external error on $\M$ vanishes, while for no external calibration the fractional change is equal to $-|\Delta \mu_{\alpha\beta}|^2$ and determined by the correlation of the bias parameters with the dark energy and cosmological parameters.

The variance of $\M$ estimated from the survey is $\sigma_{\small\M}^2 = 1/N_{\rm eff}$,
where $N_{\rm eff}$ is the effective number of independent modes measured in the shear power spectrum.
If we require the contribution to the DEFoM from bias calibration to be less than $10\%$, so that $\nu \le 0.1$, and assume that the number of effective modes measured in the shear power spectrum is $N_{\rm eff} \approx 10^{5}$,
and  $|\Delta \mu_{\M w_0}| \approx 1$, 
then the error on the multiplicative calibration needs to be less than $0.1\%$, or $\widehat{\sigma}_{\! \small \M} < 10^{-3}$. However, if the Fisher correlation coefficient is less than unity this constraint will weaken.
Similarly the fractional bias in the DEFoM from a constant  additive bias has the same form as equation (\ref{eq:DEFOMM}) with $\M \rightarrow \calA$, where $\sigma^2_{\! \small \calA} = \calF_{\small \calA\calA}^{-1} = \Delta {\overline C}$ is the inverse-weighted mean power, which implies the calibration  error on the additive bias power calibration should be $\widehat{\sigma}_{\small \calA} < 10^{-12} \,{\rm rad}^2$.

\begin{figure}
  \centering
  \vspace{-0.0cm}
   \includegraphics[width=1.08\columnwidth]{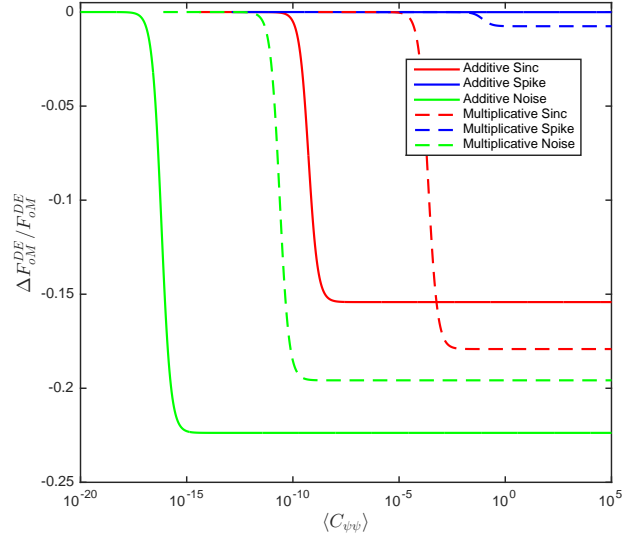}
  \caption{\em  
   The fractional decrease  in the dark energy Figure-of-Merit,  $\Delta \ln F^{\rm DE}_{\rm FoM}$, due to noise-like (green curves), sinc-like (red curves),  and spike (blue curves) bias power calibration, removal and marginalisation, as a function of the prior calibration error in the bias power amplitude. Solid lines are for additive biases, while dotted lines are for convolutions biases.  For a small external calibration  error  the change in $\Delta \ln F^{\rm DE}_{\rm FoM}$ vanishes. For larger bias error the DEFoM decreases until the data itself calibrates the bias at the cost of constraints on other cosmological parameters. 
    }
    \label{fig:4}
\end{figure}

As well as constant biases, we can investigate the more realistic cases of the spike, noise and sinc-like bias functions. 
Using the bias power functions introduced in Section 3 and illustrated in Figure \ref{fig:1} with the same parameters, we can add them together to form a multiplicative/convolved bias and us the same function as an additive bias power. We can then vary the accuracy with which we can measure the amplitude of the multiplicative/convolved and additive bias powers, to see its effect on the change in the DEFoM.
Figure \ref{fig:3} shows the fractional change in the DEFoM as a function of the variance of the external calibration for constant multiplicative and additive biases from a numerical calculation of the DEFoM.  As expected, as the external calibration accuracy decreases, the DEFoM is reduced and tends to a constant beyond $\Delta \ln F^{\rm DE}_{\rm FoM} = -0.3$, when the bias is self-calibrated by the survey. For an accurate calibration of the additive bias amplitude, we find the calibration accuracy on the multiplicative/convolution bias roughly agrees with our naive estimate, 
where a $10\%$ change, $\Delta \ln F^{\rm DE}_{\rm FoM} = -0.1$, requires a calibration error of around $\widehat{\sigma}_{\rm multi} \approx 0.07$, estimated from Figure \ref{fig:3}. However, the additive calibration error is much higher at around $\widehat{\sigma}_{\rm add} \approx 10^{-8}{\rm rad}^2$. Both of these constraints from calibration appear to be weaker than the constraints derived from requiring an algorithmic bias correction (e.g. Massey et al. 2013).

Figure \ref{fig:4} shows the change in the DEFoM due to the calibration of each individual additive and multiplicative (convolution) bias spectra for spike, noise- and sinc-like functions, as a function of the external calibration error on each bias calibration. Again, we assume only the amplitude is to be calibrated with a fixed functional form. 

The additive spike bias (solid blue line) has the smallest effect on the DEFoM, because the addition of a spike in the shear power has very little correlation with cosmological parameters. This agrees with the analysis of Kitching et al. (2015), who found a signifiant bias from an additive spike, but little increase in the $w_0, w_a$ error ellipse (see Figure 1 of Kitching et al., 2015).
However, convolution with the shear power extends this over a range of wavenumbers, and acquires  a cosmological dependence (see Figure \ref{fig:1}, left panel). Removal of this bias requires only  a modest accuracy, $\widehat{\sigma}_{\rm spike, add} \approx 1$.
As this cosmological dependence still does not mimic the effects on the true shear power, self-calibration with a shear survey works well, resulting in only a small reduction in the DEFoM. 

The noise bias (green lines) has the highest calibration requirements, with the additive bias (solid green) requiring a high accuracy of $\widehat{\sigma}_{\rm noise,add} < 10^{-8}$ for removal, while self-calibration leaves the largest reduction in the DEFoM. 
This can be understood from the noise power having greatest effect at high wavenumber, where the dark energy equation of state will have greatest effect. The multiplicative/convolution bias also requires a high accuracy to calibrate, as the resulting convolution is still noise-like and has a cosmological dependence.

The sinc-function bias power is like the noise-like bias power, with an effect on the DEFoM between the noise and spike bias power,  with the oscillations dampening the bias power at high wavenumber, which  de-correlates the bias and cosmological parameters. Again, the additive sinc bias requires more accurate calibration than the multiplicative/convolution bias.
Finally, we find the accuracy of the calibration scales with the amplitude of the bias, such that numerically we find $\widehat{C}_{\psi\psi} (\psi) \propto \psi^{-1}$,  for all biases. 

In summary, we find the calibration and removal of noise-like biases has the greatest impact on dark energy studies, followed by sinc-like biases. The effect of calibration and removal of spikes in the shear power spectrum has  the least effect. Caveats to this study are that we consider  only calibration of the amplitude of these bias effects. In detail, for the spike and sinc bias functions, we would also want to calibrate the  scale at which the bias occurs, while  the functional forms of the systematic power may require many more parameters to describe.

\section{Optimising bias calibration }

Since we can expect the calibration of any shear bias has a cost, either in the collection of external calibration data or the generation of realistic simulations, is is useful to have a guide for where to optimally allocate resources in investigating sources of bias, their calibration and removal. 
Here we shall assume that any bias can be modelled through a simulation of the experiment, and that the accuracy on the measurement of any bias is limited only by the number of simulations that can be generated. A similar calculation can be done if the cost of  the calibration arises from collecting external data. We assume these biases are independent, and work to first order in the external calibration error, $\widehat{\sigma}^2_\psi$, so that the effect on the DEFoM can be written $\Delta \ln {\rm F_{oM}^{DE}} =  -\Phi_\psi \widehat{\sigma}^2_\psi$, where $\Phi_\psi =  \calF^M_{\psi w} \calF^{-1}_{ww'} \calF^M_{w' \psi}$ and there is no summation over repeated $\psi$ in this last expression.
 If we further assume that each bias requires its own set of simulations for calibration, the total number of simulations needed to calibrate all biases with accuracy an of $\widehat{\sigma}^2_\psi$ scales as
\be
	N_S = \sum_\psi \frac{\alpha_\psi}{\widehat{\sigma}^2_\psi},
\ee
where $\alpha_\psi$ is a parameter which normalises the number of simulations needed to calibrate each bias and depends on the properties of the bias. Our aim is to minimise the number of simulations needed for calibration, with the constraint that we do not exceed the desired fractional change in the DEFoM, $\nu= |\Delta \ln {\rm F_{oM}^{DE}}|$. We can calculate this by minimising the merit function,
\be
	S = N_S + \lambda \nu,
\ee
with respect to $\widehat{\sigma}^2_\psi$, where $\lambda$ is a Lagrangian multiplier to constrain the DEFoM. Minimising this with respect to the measured external calibration error, $\widehat{\sigma}^2_\psi$, taking note that $\nu$ is the absolute value of the fractional change in the DEFoM, and using the identity $|x|=\sqrt{x^2}$, we find
\be
	\widehat{\sigma}^2_\psi = \left|\frac{\alpha_\psi}{\lambda \Phi_\psi} \right|^{1/2}.
\ee
Using the constraint on the DEFoM we can replace the Lagrangian multiplier, $\lambda$, with $\nu$, 
to find that the minimum number of simulations is
\be
	N_S = \frac{1}{\nu} 
	\left(\sum_{\psi'}   \alpha_{\psi' } \left|\frac{\Phi_{\psi'}}{\alpha_{\psi'}} \right|^{1/2} \right)
	\left|\sum_{\psi''}   \Phi_{\psi''} \left|\frac{\alpha_{\psi''}}{\Phi_{\psi''}} \right|^{1/2} \right|,
\ee
which yields  the error on the bias calibration,
\be
	\widehat{\sigma}^2_\psi  = \nu
	\left| \frac{\alpha_\psi}{\Phi_\psi} \right|^{1/2}
	\left| \left(\sum_{\psi'}   \Phi_{\psi' } \left|\frac{\alpha_{\psi'}}{\Phi_{\psi'}} \right|^{1/2} \right)\right|^{-1}.
\ee

As an example, let us assume that each simulation is of the entire survey, so that we can the calibrate a bias with an error $\sigma_\psi$. To reach the required calibration error, $\widehat{\sigma}_\psi$, we need $\sigma_\psi^2/\widehat{\sigma}^2_\psi$ simulations per bias parameter, and the total number of simulations is $N_S = \sum_\psi \sigma_\psi^2/\widehat{\sigma}^2_\psi$. Hence, $\alpha_\psi = \sigma_\psi^2$.
For a single dark energy parameter, $w=w_0$, a single cosmological parameter, $\theta$, and summing over all bias parameters, the fractional change in the DEFoM is 
\be
	\Delta \ln {\rm F_{oM}^{DE}}=-\sum_\psi |\Delta \mu_{w \psi}|^2 \frac{ \widehat{\sigma}^2_\psi}{\sigma^2_\psi},
\ee
 so that $\Phi_\psi =|\Delta \mu_{w \psi}|^2 /\sigma^2_\psi $. 
The number of simulations needed for calibration is then
\be
	N_S = \frac{1}{\nu}  \left(\sum_{\psi}  |\Delta \mu_{w \psi}|  \right)^2.
\ee
If  the correlation between the bias and $w$ is zero, no simulations are required. When the correlation between the bias and $w$ is of order unity and the number of bias parameters is $N_{\rm bias}$, then the number of simulations we require is of order $N_S \approx N^2_{\rm bias}/\nu$.
For the simple example of the six bias normalisation parameters used in this paper, $N_{\rm bias} = 6$, and for $\nu =0.1$, we expect $N_S \approx 360$. With more detailed numerical studies using our Fisher Matrix formalism, and the spike, noise and sinc functional forms, we find $N_S \approx 100$. Given the quadratic scaling with the number of bias parameters, we can expect this number to rise rapidly. If we have 100 calibration parameters to measure we may need $N_S \approx 10^6$ simulations. However, these simulations may have the same underling simulation, adding on the effect of each systematic.

The resulting variance on the measured calibration is 
\be
	\widehat{\sigma}^2_\psi =\frac{\nu \sigma^2_\psi}{|\Delta \mu_{w \psi}|} 
				\left(\sum_{\psi'}  |\Delta \mu_{w \psi'}|  \right)^{-1} ,
\ee
so that the calibration error is inversely proportional to the marginalised Fisher correlation coefficient, $\Delta \mu_{w \psi}$. Again, if this is of order unity the external calibration error is $\widehat{\sigma}_\psi \approx  \sigma_\psi \sqrt{\nu/N_{\rm bias}}$.

\section{Summary and Conclusions}

In this paper we have extended the analysis of cosmic shear  to include the effect of spatially varying image distortions on the sky; investigated the effect of calibration of biases from external data or simulations and internally from the same survey on cosmological parameter estimation and the dark energy Figure-of-Merit; and shown how to minimise the size of the calibration set for a given impact. 
Spatially varying image distortions convolve the shear signal in Fourier space, and the shear power spectrum,  mixing even-parity convergence and odd-parity $\beta$-mode signals. We have found analytic solutions for the biased shear-convergence and $\beta$-mode power from convolution with spike, noise and sinc-like bias power spectra, which can be generated in realistic cosmic shear surveys. In all cases we have studied, the bias power is equally distributed between convergence and $\beta$-mode power.  We find that a scale-dependent multiplicative bias power spectrum model, which has commonly been used in previous studies, is not an accurate approximation.

Convolution and additive biases can be removed from the signal by calibration to external data, or from simulations of the effect of the bias, or by allowing the bias to be fit simultaneously to the data. In such a scenario the absolute value of the bias is unimportant since it will be removed and marginalised over, but the uncertainty in the calibration will propagate into the measurement of cosmological and dark energy parameters. We have carried out an analysis to show how removal and marginalisation of the bias, using calibration data and self-calibration from the data itself, will propagate into cosmological parameter estimation and then into the dark energy Figure-of-Merit. We have applied this to archetypal functions forms for the bias power, spike, noise and sinc-like functions, and show how each individually, and in combination, affect the dark energy Figure of Merit. We find that calibration and removal of the noise-like bias functions, which affects the largest range of scales, has the greatest affect on the  DEFoM, followed by the sinc-like function, which contains an oscillatory cut-off at small scales, while the spike bias has the least effect, covering the smallest range of scales. Overall, a calibration approach appears to require less stringent constraints on bias errors than the algorithmic corrections of the bias.

We have also carried out an optimisation of the required calibration error, in order to minimise the number of simulations needed to measure the calibration for a fixed deterioration of the DEFoM. This calculation could also be used to minimise the external data required for calibration.

Finally, our method is general enough that we can extend the formalism to allow the study of bias, bias-removal and the effect of calibration error in the nonlinear matter power spectra, baryonic effects on the matter power spectra, photometric redshift calibration, intrinsic alignment calibration, and indeed any effect in the measurement which can be corrected for by calibration. As the formalism is a Pseudo-Cls approach  it can account for the effects of the survey window function on the shear power spectrum. This enables the investigation of the effect removing these biases on dark energy and modified gravity experiments.

\section*{Acknowledgements}

ANT thanks the Royal Society for the support of a Wolfson Research Merit Award while TDK is supported by a Royal Society University Research Fellowship. ANT also acknowledges the support of the UK Space Agency and an STFC Consolidated Grant. We thank Mark Cropper and Tim Schrabback for useful discussions.

{}

\appendix

\onecolumn

\section{Fourier modes of biased shear $\kappa$ and $\beta$ fields}
The change in the Fourier modes of the convergence, $\Delta \kappa = \widehat{\kappa} - \kappa$, and $\beta$-modes $\Delta \beta = \widehat{\beta} -\beta$, transformed from the measured shear in equation (\ref{eq:2}), with   a spatially varying spin-2 multiplicative distortion  of the shear signal, which we write as $m=m_{1} + i m_{2}$, is
\ba
	\Delta \kappa(\ellb) \!\!\! &=&  \!\!\!\!\!
		\int \! \frac{d^2 \ell'}{(2\pi)^2}\, 
		\Big(m_{1}(\ellb-\ellb') \left[\kappa(\ellb') \cos 2 \varphi_{\ell\ell'}   -
		  \beta(\ellb')	\sin 2 \varphi_{\ell\ell'} \right]  
		    -
		  m_{2}(\ellb-\ellb')
		 \left[\beta(\ellb') \cos 2 \varphi_{\ell\ell'} +
		  \kappa(\ellb')	\sin 2 \varphi_{\ell\ell'} \right] \Big)
		  + c_\kappa(\ellb), \\
	\Delta \beta(\ellb) \!\!\! &=& \!\!\!\!\!
		 \int \! \frac{d^2 \ell'}{(2\pi)^2}\, 
		 \Big(  m_{1}(\ellb-\ellb') \left[\beta(\ellb') \cos 2 \varphi_{\ell\ell'} +
		 \kappa(\ellb') \sin 2 \varphi_{\ell\ell'} 	\right]
		  				   +
		  m_{2}(\ellb-\ellb')
		 \left[\kappa(\ellb') \cos 2 \varphi_{\ell\ell'}  - \beta(\ellb')	\sin 2 \varphi_{\ell\ell'}    \right] \Big)
			 + c_\beta(\ellb) .
\ea
The first terms are the same as for a scalar multiplicative distortion, when the phase angle is $\varphi_m = 0$. The second term in the outer brackets of the integrand is due to the spin-2 local phase change of the multiplicative bias mixing in the orthogonal component of the bias.

\section{Correlations of the observed Fourier modes}
\label{Correlations of the observed Fourier modes}

The correlation of the Fourier modes of the measured convergence field, assuming no intrinsic $\beta$-modes, $C^{\beta\beta}_\ell=0$, and for a scale-dependent, scalar multiplicative bias, $m_0(\ell)$ where each bias is broken down into spatially constant, deterministic spatial variations and random variations, $m_0(\ellb) = b_m (\ellb) + \delta m(\ellb) $ where $b_m(\ellb) = b_m \delta_D(\ellb) + \Delta m(\ellb)$,
is given by
\ba
	 \label{eq:conv_FFT_corr}
	\lgl \widehat{\kappa}(\ellb) \widehat{\kappa}^*\!(\ellb') \rgl &=&
	(2 \pi)^2 C^{\kappa\kappa}_\ell \delta_D(\llb-\llb')
	 +
	 \left(b_{m}(\ellb-\ellb') C^{\kappa\kappa}\!(\ell) 
	 +
 	 b_{m}(\ellb'-\ellb)  C^{\kappa\kappa}\!(\ell') \right) \cos 2 \varphi_{\ell\ell'}
	  +
	 b_{c_\kappa}(\ellb) b^*_{c_\kappa}(\ellb') 
	 + (2 \pi)^2 \calC_{c_\kappa}(\ell)  \delta_D(\ellb-\ellb')
	 \nn
	 & & 
	 +
	 \int \! \frac{d^2 \ell''}{(2\pi)^2} \Big(b_m(\ellb-\ellb'')b^*_m(\ellb'-\ellb'') 
	 +
	 (2\pi)^2 \calC_{m}(\ellb-\ellb'') \delta_D(\ellb-\ellb') \Big) C^{\kappa\kappa}\!(\ell'')
	 \cos 2 \varphi_{\ell \ell''}\cos 2 \varphi_{\ell'\ell''} ,
\ea
\ba
	 	\lgl \widehat{\beta}(\ellb) \widehat{\beta}^*\!(\ellb') \rgl &=&
	 b_{c_\beta}(\ellb) b^*_{c_\beta}(\ellb') 
	 + (2 \pi)^2 \calC_{c_\beta}(\ell)  \delta_D(\ellb-\ellb')
	 \nn
	 & & 
	 +
	 \int \! \frac{d^2 \ell''}{(2\pi)^2} \Big(b_m(\ellb-\ellb'')b^*_m(\ellb'-\ellb'') 
	 +
	 (2\pi)^2 \calC_{m}(\ellb-\ellb'') \delta_D(\ellb-\ellb') \Big) C^{\kappa\kappa}\!(\ell'') 
	 \sin 2 \varphi_{\ell \ell''}\sin 2 \varphi_{\ell'\ell''},
\ea
and
\ba
	 \label{eq:conv_FFT_corr2}
  	\lgl \widehat{\kappa}(\ellb) \widehat{\beta}^*\!(\ellb') \rgl &=&
	 b_{m}(\ellb-\ellb') C^{\kappa\kappa}\!(\ell)  \sin 2 \varphi_{\ell\ell'}
	  +
	 b_{c_\kappa}(\ellb) b^*_{c_\beta}(\ellb') 
	 \nn
	 & & 
	 +
	 \int \! \frac{d^2 \ell''}{(2\pi)^2} \Big(b_m(\ellb-\ellb'')b^*_m(\ellb'-\ellb'') 
	 +
	 (2\pi)^2 \calC_{m}(\ellb-\ellb'') \delta_D(\ellb-\ellb') \Big) C^{\kappa\kappa}\!(\ell'') 
	 \cos 2 \varphi_{\ell \ell''}\sin 2 \varphi_{\ell'\ell''}. 
\ea
From these expressions we can take $\ellb=\ellb'$, and use the finite-field approximation $\delta_D(\0) = A/(2 \pi)^2$, to yield the convergence, given by equations (\ref{eq:obs_conv}) and (\ref{eq:kernel}) and the $\beta$-mode power given in Appendix \ref{B-mode power spectra}.

\section{$\beta$-mode power spectra}
\label{B-mode power spectra}

The measured $\beta$-mode power spectrum is
\be
	 \widehat{C}^{\beta\beta}\!(\ell) =  \frac{1}{A}	\lgl |\widehat{\beta}(\ellb)|^2 \rgl =
	 \int \! \frac{d^2 \ell'}{(2\pi)^2} \Big(\frac{1}{A}|b_m(\ellb-\ellb')|^2
	 +
	  \calC_{m}(\ellb-\ellb') \Big) C^{\kappa\kappa}\!(\ell')
	 \sin^2 2 \varphi_{\ell \ell'} +  \calA^{\beta\beta}\!(\ell),
\ee
where 
\be
 \calA^{\beta\beta}\!(\ell) = 	\frac{1}{A} |b_{c_\beta}(\ellb) |^2 
	 +  \calC_{c_\beta}(\ell) .
\ee
If $b_m$ is a constant the bias factors drop out and the measured $\beta$ power becomes 
\be
	 \widehat{C}^{\beta\beta}\!(\ell) =  
	 \int \! \frac{d^2 \ell'}{(2\pi)^2} 
	  \calC_{m}(\ellb-\ellb')  C^{\kappa\kappa}\!(\ell') 
	 \sin^2 2 \varphi_{\ell \ell'} +  \calA^{\beta\beta}\!(\ell).
\ee
A constant noise-power bias, $\calC_m(\ell) = \calC_{\! N}$, will lead  to a  $\beta$-mode power of $\widehat{C}^{\beta\beta}\!(\ell) = \calC_{\! N} \sigma_\kappa^2 /2$.
If we assume a slowly varying scale dependent shear power spectrum reduces to
\be
\label{eq:Cbb}
		 \widehat{C}^{\beta\beta}\!(\ell) =   \frac{1}{2} \sigma^2_m C^{\kappa\kappa}\!(\ell) +  \calA^{\beta\beta}\!(\ell).
\ee
Hence we see that the mixing of Fourier modes is required to transform from convergence to $\beta$-mode  power, and that the assumption of a slowly varying shear power spectrum evenly distributes the measured multiplicative power between the convergence and $\beta$ modes.

\end{document}